\documentclass{article}
\PassOptionsToPackage{hyphens}{url}   %

\usepackage{iclr2027_conference,times}
\iclrfinalcopy %
\usepackage{graphicx}
\usepackage{flafter} %
\usepackage{placeins} %
\usepackage{float}
\floatstyle{ruled}
\newfloat{algorithm}{tbp}{loa}
\floatname{algorithm}{Algorithm}
\newcounter{algln}
\newif\ifalgnum \algnumtrue
\newlength\alglnw 
\newlength\alglns 
\newlength\algind 
\newcommand\AlgSetup{\setlength\parindent{0pt}\setlength\parskip{0pt}\rightskip=0pt plus 1fil\relax\hyphenpenalty=10000\exhyphenpenalty=10000\relax}
\newcommand\Ln[2]{\par\noindent
  \ifalgnum\refstepcounter{algln}\hangindent=\dimexpr\alglnw+\alglns+#1\algind\relax
  \else\hangindent=\dimexpr#1\algind\relax\fi\hangafter=1
  \ifalgnum\makebox[\alglnw][r]{\footnotesize\thealgln}\hspace{\alglns}\fi
  \hspace*{#1\algind}#2\par}
\newcommand\kw[1]{\textbf{#1}}
\newcommand\Proc[1]{\textsc{#1}}
\usepackage{amsmath,amssymb}
\usepackage{booktabs}
\usepackage{array,tabularx}
\usepackage{longtable}
\usepackage{xcolor}
\usepackage[hidelinks]{hyperref}
\usepackage{xspace}
\usepackage{enumitem}

\makeatletter
\newcommand{\hidecomment}{\@bsphack\@esphack}
\makeatother

        \newcommand{\todo}[1]{\hidecomment}
        \newcommand{\gf}[1]{\hidecomment}
        \newcommand{\nicolasc}[1]{\hidecomment}
        \newcommand{\taro}[1]{\hidecomment}
        \newcommand{\ch}[1]{\hidecomment}
        \newcommand{\review}[1]{\hidecomment}

\definecolor{plotdark}{HTML}{385CB1}
\definecolor{plotdark}{RGB}{46,82,237}
\definecolor{plotlight}{HTML}{98B3E7}
\newcommand{\taubench}{$\tau^3$-bench\xspace}

\newcommand{\mypara}[1]{\noindent \textbf{#1.}}

\newcommand{\name}{verifiable agent insurance\xspace}

\newcommand{\myparagraph}[1]{\noindent \textbf{#1}.}

\definecolor{covdetected}{HTML}{2B5DB7}
\definecolor{covadded}{HTML}{91B4EC}
\definecolor{covsubjective}{HTML}{D6D5D1}

\title{Holding Agents Accountable\\with Verifiable Agent Insurance}
\author{Cheng Hua, Taro Tsuchiya, Nicolas Christin \& Giulia Fanti \\
Carnegie Mellon University \\
\texttt{\{chua2,ttsuchiy,nicolasc,gfanti\}@andrew.cmu.edu}}

\begin{document}
\maketitle
\lhead{Preprint} %

\begin{abstract}
While AI agents are useful for many tasks, they regularly misbehave in destructive ways, such as deleting important data. 
While prior work has focused on preventing such misbehavior, we consider how to recover \emph{after} an agent misbehaves. 
We propose \name (VAI), a framework for agent providers to commit financially to \emph{promises}, a list of programmatically-specified policies that an agent is meant to follow.
These promises are embedded in a smart contract on an immutable ledger (e.g., a blockchain) to prevent tampering, along with a monetary escrow. 
A user who detects a violation of committed promises
can prompt a verifier to review agent execution traces, which are 
logged through the agent harness’s shared tool-execution path, without instrumenting each tool separately.
Upon verification of the violation, the agent provider's escrow funds are paid to the reporting user.
We instantiate VAI as an end-to-end implementation
on an Ethereum layer-2 network, and evaluate its coverage on three prominent agent benchmarks, two of which we re-annotated to obtain task-level policy-compliance labels.
Currently, many agents do not emit user authorizations and security-relevant information as explicit, structured data objects---rather, they embed the data needed to verify policy violations in natural language exchanges. Even so, VAI manages to detect violations in 63.6\% of runs containing violations across three benchmarks. By augmenting agents to explicitly emit such objects, violation detection rises to 79.8\% of runs containing violations.
Our findings 
suggest that for improved auditability, agents should be co-designed with their security policies to emit explicit, structured records of user authorization and security-relevant actions.
\end{abstract}

\section{Introduction}
\label{sec:introduction}

Modern AI agents interact with the world by executing payments, modifying databases, and communicating in critical applications, to name a few.
Agents are now used for both organizational and personal tasks~\citep{johnston2026shift}, but they can also cause serious harm. 
Prominent examples of AI agents behaving in unintended and harmful ways abound, including a Replit coding agent that deleted a production database~\citep{replit2025register},
an OpenClaw agent that mistakenly deleted emails belonging to an alignment director at Meta Superintelligence Labs~\citep{yue2026fastcompany}, 
 the coordinated attack on the Hugging Face platform by OpenAI agents escaping containment~\citep{openai2026hfroadahead}, and more \citep{grok2026neuraltrust,anthropic2026cyberevalincidents}. 

The AI community currently addresses the risk of agent misbehavior by attempting to \emph{prevent} misbehavior via safeguards.  
Some safeguards are system-level restrictions: OpenAI and Anthropic use approval requirements and sandboxes to restrict agents' actions and access to files and networks~\citep{openai2026codexsafety,dworken2025sandboxing}. 
Other safeguards evaluate more nuanced context and match it to behavioral policies; researchers have proposed systems that enforce agent compliance with behavioral policies, sometimes assisted by large language models (LLMs)~\citep{shi2025progent,debenedetti2025camel,li2026drift}.  
These measures are all important safeguards, but the attack surface for agents is vast, and preventing every failure is challenging.

Our paper asks a different question: what happens once an agent misbehaves? 
This question has serious implications for the growing agent provider industry, particularly as agents are being sold to organizations with strict compliance and financial requirements \citep{palantir,accenture}.
When agents cause harm today, 
user recourse depends on human processes. 
For example, OpenAI's May 2025 business terms and Anthropic's June 2025 commercial terms cap liability at twelve months of fees, subject to exceptions~\citep{openai2025terms,anthropic2025terms}.
After the Replit incident, its CEO publicly promised the affected user a refund~\citep{council2025replit}, but only at the CEO's goodwill.

Existing forms of recourse for agent misbehavior have three key problems: 
(1) They are slow and subjective, requiring human review. 
(2) Compensation is not necessarily commensurate with the user's losses, as in the case of OpenAI's and Anthropic's policies. 
(3) A claim may depend on logs held by the provider, leaving the user to obtain evidence from the party facing the claim. 
The latter point is particularly problematic, given that misbehaving agents may try to alter their own records. 
METR's investigation of the Hugging Face breach found agents researching ways to tamper with their transcripts, and altering what their tool calls returned in about 7\% of reviewed transcripts~\citep{metr2026incident}. 
During training, OpenAI also observed an agent write a task summary proposing to invent missing historical data and conceal that fact from the user~\citep{openai2026summarydeception}.
An agent's own account is therefore insufficient evidence for adjudication.

We tackle these problems by proposing a new framework for agent accountability: in \emph{\name} (VAI)
agent providers register a set of \textit{promises}, each asserting that the agent will not violate a given policy. 
Promises are backed by a deposit from the agent provider, which is escrowed on an immutable ledger (e.g., a public blockchain). 
During agent execution, agent actions (e.g., tool calls) are signed and sent to the user and committed periodically as hashes to the ledger, while the underlying execution traces remain with the agent provider until verification.
A user who detects a violation of a promise can file a challenge.
If verified,
the escrow smart contract slashes (pays) the specified amount from the provider's reserve.
We envision agent providers being motivated to implement \name as a guarantee of the trustworthiness of their agents, potentially as a precursor to procurement.

\paragraph{Contributions}
Our contributions are threefold. 
\begin{itemize}[leftmargin=*]
    \item \textbf{Novel accountability framework:} We propose a \emph{\name} (VAI) accountability framework for AI agents. This framework binds promises regarding agent behavior and agent execution traces to escrowed financial reserves, with automatic settlement in case of violations. 
    Agents typically alter the state of real-world systems through tool calls, and  
    route tool calls through a shared dispatch point, whose instrumentation lets us record tool names, arguments, and reported results without modifying each tool individually. 
    \item \textbf{Instantiation and evaluation:} We instantiate \name with an SDK that logs tool-calling data through adapters for different agent harnesses.
    Our end-to-end implementation runs on Base mainnet, an Ethereum layer-2 network.\footnote{\url{https://github.com/ChengHua926/vai-artifact}}
    Across AgentDojo, $\tau^3$-bench, and ClawsBench, VAI detects at least one violation in 63.6\% of runs containing violations.
    Many missed violations depend on information that is not exposed as structured objects.
    Through updating the harness to structurally emit data objects, we show that coverage can be extended to 79.8\% of these runs.
    At the same time, across the three benchmarks, the pooled false-positive rate is 6.7\% with the original harness and 7.4\% with the updated harness.
    \item \textbf{Dataset curation:} Existing agent misbehavior benchmarks (including the ones we used) often annotate data at a level of granularity that is too coarse to understand precisely \emph{how} an agent fails. For example, the $\tau^3$-bench version we evaluate checks a database state after agent interactions, as well as pre-specified, task-specific assertions and communication requirements~\citep{yao2024tau,sierra2026tau3}.
    These checks do not establish compliance with every policy obligation or identify all violated rules. For our evaluation, we therefore annotate 388 agent runs from $\tau^3$-bench and ClawsBench with task-level policy-compliance labels.
    We release these annotations as an independent contribution of the work.\footnote{\url{https://github.com/ChengHua926/vai-annotations}}
\end{itemize}

Overall, \name is only a first step towards truly accountable agents. 
Our instantiation is currently limited to violations that can be established programmatically from recorded traces.
At the same time, the AI community needs better methods for holding agents accountable during and after failures, and the \name framework could be a starting point, both for insuring current agents and for encouraging harness developers to more systematically record agent actions in a way that can be effectively monitored and punished in case of malfeasance. 

\section{Setup}
\label{sec:gap}
\label{sec:overview}

\myparagraph{Stakeholders} We consider four key stakeholders. 
(1) The \emph{agent} is a computer program that consists of a reasoning engine (typically implemented by an LLM) and a harness, which manages the interface with the user, calls to the reasoning engine, temporary storage, and tool calls.
(2) The \emph{agent provider} operates and develops the agent; this party typically is responsible for testing and securing the agent.
The provider has a set of natural-language policies for agent behavior and is responsible for ensuring that the agent adheres to these policies.
(3) The \emph{user} interacts with the agent by inputting natural language prompts and responding to queries from the agent. 
In our model, the user is responsible for identifying agent violations and submitting a claim. 
(4) Finally, the \emph{verifier} is responsible for verifying user claims of misbehavior. 

\myparagraph{Policy violations} We categorize policy violations into two types: \emph{objective} and \emph{subjective}. Objective policies can be specified as a formal rule that can be verified.  For instance ``\texttt{never send money to account 183509}." Subjective rules instead require interpretation, such as ``\texttt{never send money to people I don't like}." 
The division of policies into objective and subjective rules can depend on the agent in question, 
e.g., if consent is collected implicitly from user prompts. Some systems collect explicit approval decisions that are available to the harness~\citep{anthropic2026permissions,openai2026codexapprovals}. 
We focus on objective violations.

\myparagraph{Assumptions and capabilities} We assume the existence of an immutable ledger (such as a permissionless blockchain) that is live and correct.
We call an agent's interactions with a user a \emph{session}, $s$. Relevant tool calls and permission decisions form an execution \emph{trace}, $T$.

\noindent \emph{Provider.} We assume the provider 
correctly integrates our unmodified software development kit (SDK) at every tool-dispatch and permission path to
collect and store them as execution traces $T$. 
Those traces $T$ contain true tool invocations, arguments, and outcomes (e.g., changes to the filesystem or state of the machine on which an agent is running).
The agent cannot fabricate user-approval records.
While we assume faithful SDK integration, we do not assume that the provider will preserve or supply those records honestly once the trace has been generated.

\noindent \emph{Verifier.}
We assume the verifier follows protocol. Our reference implementation relies on an independent verifier for correct and timely adjudication.
In practice, this could be achieved by assigning the verification role to a neutral, independent, trusted third party (as in our implementation) or by requiring both the provider and verifier to cryptographically prove they executed their tasks correctly. 
In the latter case, verification could even be run by the agent provider itself. 

\noindent \emph{Agent.} The agent may violate policies and alter or delete execution records to which it has write access. We assume it cannot modify its harness or SDK integration, bypass the recording of covered actions, or fabricate user approvals. 

\noindent \emph{User.} The user may submit invalid claims or deliberately induce violations to collect payouts.

\myparagraph{Problem statement} Our goal is to define a framework such that if an agent commits an \emph{objective} policy violation, any party involved in the session and registered as its claimant can submit a claim. If the agent  violated the policy, 
the agent provider should deterministically lose funds. 
The system should be robust to misbehavior by the agent, agent provider, and user, within our threat model.

\myparagraph{Non-goals} We do not address attacks that compromise the recording of traces, 
including harness modification or delegation to uninstrumented agents. 
Our focus is financial recourse for policy violations---whether accidental or deliberate---that can be established from faithfully-recorded execution traces.
Extending our framework to a stronger adversarial model is an important direction, which we leave for future work.

\section{Verifiable Agent Insurance Design}
\label{sec:design}
In this section, we explain how (1) agent providers register promises and deposit collateral, (2) our SDK records agent traces, and (3) users submit violations and verifiers adjudicate claims
(Figure~\ref{fig:overview}).
Pseudocode for \name is provided in Appendix \ref{app:pseudocode}.

\begin{figure}[!htbp]
    \centering
    \includegraphics[width=.75\textwidth]{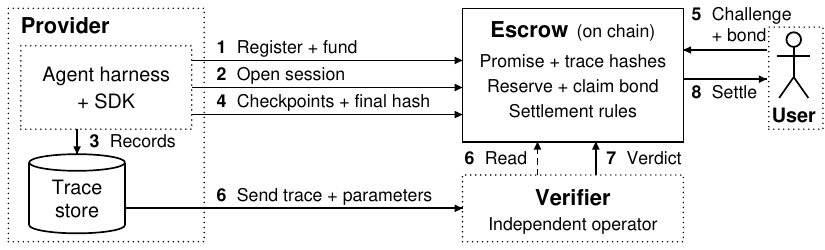}
    \caption{VAI protocol overview.
    The provider registers a promise and deposits its reserve (1), then opens a session for the user (2). The SDK sends execution records to the off-chain trace store (3); the provider commits a checkpoint and final trace hashes on-chain (4). The user files a challenge with a bond (5). The provider sends the trace and parameters to the verifier, which reads the commitments (6), checks the evidence and, when it matches, evaluates the promise before submitting a verdict (7). The escrow applies the settlement rules and pays eligible claims from the available reserve (8). Dotted boxes identify stakeholders. }
    \label{fig:overview}
\end{figure}

\FloatBarrier
\subsection{Specifying and funding a promise}
\label{sec:design:promises}
Providers must specify their promises and deposit collateral before users begin using their agents. Consider a provider promising that its email assistant will email only recipients on a specified list. Each promise has a predicate $\phi$, configuration parameters $\theta$, a payout $p$, and an associated, time-dependent reserve $R_t$:
\begin{itemize}[leftmargin=*]
    \item $\phi(T;\theta)$ is a predicate to check whether the recorded trace $T$ violates a rule. For example, $\phi$ may check whether any recorded email features a recipient outside the allowed list.
    \item $\theta$ is a set of parameters to configure the predicate. The parameters in our example include: the tool type (e.g., \texttt{send\_email}) and its argument of interest (e.g., \texttt{to} field), and an allowed list (e.g., $[$\nolinkurl{alice@example.com}, ...$]$). 
    \item $p$ is a payout paid to the user when a claim qualifies for payment. For example, $p=0.1$ ETH for sending emails to unknown recipients.
    \item $R_t$ is a reserve available at time $t$ to fund the promise payouts across all sessions. For instance, providers can hold an initial reserve of $R_0=1$ ETH. 
    When a user files a challenge, the escrow sets aside $p$ from $R_t$, pays it to the user if the violation is proven, and otherwise returns it to the reserve. Claims are funded in filing order, so a claim filed while $R_t<p$ waits until deposits or returned set-asides cover it (Appendix~\ref{app:design:funding}).
    As a starting point, we assume each promise has its own payout.
\end{itemize}

The predicate evaluates the trace $T$ under parameters $\theta$: $\phi(T;\theta)$. 
In our example, if $\theta$ specifies that only \nolinkurl{{alice@example.com}} is allowed, $\phi$ returns \texttt{satisfied} if $T$ only contains emails to \nolinkurl{alice@example.com}, and \texttt{violated} otherwise.
The same trace and parameters always produce the same verdict.
Providers can use natural-language policies or regulations (e.g., OFAC regulations) to either derive promises manually or use LLMs to facilitate promise specifications, as explored in prior work~\citep{wang2026agentspec,kholkar2025policyasprompt}.

We provide a catalog of reusable predicates with parameters.
Providers select a predicate and configure its tools, permitted targets, or limits. Appendix~\ref{app:design:predicates} describes these parameters.

We implement five example predicates: 1) \textbf{authorization:} no destructive actions without permission; 2) \textbf{scope:} no actions beyond an allowed scope,
3) \textbf{payment:} no payments to merchants outside an approved list or above a per-payment cap; 4) \textbf{data exfiltration:} no data sent to recipients outside a specified list; and 5) \textbf{action limits:} upper bounds on action counts, cumulative amounts, or repeated calls beyond their limits.
Providers can configure these predicates to express different promises or add predicates for other rules.

\paragraph{Collateralization}
When registering a promise, the provider specifies $p>0$ and deposits an initial reserve $R_0\geq p$ with the escrow smart contract.
Appendix~\ref{app:design:funding} describes how providers fund, replenish, and withdraw the reserve.
Importantly, we prevent providers from changing the promise after registration by recording the payout $p$ and hashes of the predicate and parameters $\theta$ in the escrow smart contract on the blockchain.
Verifiers refer to these hashes as the registered rule and parameters. 
Changing the predicate or parameters would require registering a new promise.

\FloatBarrier
\subsection{Recording and committing evidence}
\label{sec:design:evidence}
\label{sec:properties}
Once a provider has committed a promise, they may deploy many \emph{instantiations} of agents, each serving different users. 
We must track each such agent's actions to evaluate promise violations, while trying to avoid cumbersome per-tool implementations or integrations. 
Critically, major agent frameworks~\citep{anthropic2026hooks,openai2026lifecycle,vercel2026hooks} 
provide hooks for observing tool calls and their results.
The provider can use these hooks to integrate our SDK, which records each call's tool name, arguments, and reported result, within each user session.
Due to this design choice, in practice, VAI can check only what these records expose.

Some promises require evidence that an action had prior permission.
The harness may ask the user for approval or apply an existing permission rule, as modern agent harnesses already do~\citep{anthropic2026hooks,openai2026hooks}.
The SDK records the decision directly from the harness, including who or what made it and the specific call and arguments it covers.
The email assistant example from Section~\ref{sec:design:promises} may also promise to obtain permission before each send. 
The following excerpt shows approval for sending an email to Alice and the subsequent call. 
We omit some fields for brevity.
\par\noindent\begin{minipage}{\linewidth}
\footnotesize
\begin{verbatim}
[
 {"seq": 1, "tool": "user_authorization",
  "args": {"action_id": "mail-17", "tool": "send_email",
           "decision": "allow", "authority": "human",
           "scope": "once", "args_hash": "0x7c2303dc...8d7a2058"}},
 {"seq": 2, "tool": "send_email",
  "args": {"to": ["alice@example.com"], "subject": "Project update",
           "body": "The report is ready."},
  "metadata": {"action_id": "mail-17"}, "result": {"status": "sent"}}
]
\end{verbatim}
\end{minipage}
\par
The verifier makes sure the user approval comes first and authorizes that exact session, action identifier, tool, and inputs.
In this example, the approval is for this specific tool call for sending an email and does not authorize a different recipient, changed message, or another call.

\mypara{Trace logging}
The provider stores execution traces off-chain because they may contain private user data.
To defend against alterations from the provider or agent, our system creates checkpoints $(s,n,h_n)$, each containing the session identifier $s$, the record count $n$,
and $h_n=H(T_{1:n})$, a Keccak-256 hash of the first $n$ records of $T$. %
After every record, the SDK signs $(s,n,h_n)$ with a signing key generated for that session and sends it to the user.
Any change to the records covered by the user's newest signed checkpoint is then detectable.
The SDK also periodically commits checkpoints on chain, which protects users who do not keep signed checkpoints.
To limit transaction fees, the SDK batches records, attempting a checkpoint after $k$ new records or when the oldest uncommitted record reaches $r$ seconds.
Section~\ref{sec:eval:practicality} measures the costs and delays of signed checkpoints and of different on-chain checkpoint policies (e.g., $k=10$, $r=30$), Appendix~\ref{app:implementation:signed} details signed checkpoints, and Appendix~\ref{app:implementation:recording} describes how our system retries in case of failure.
At session close, our system commits $H(T)$, a hash of the complete stored trace.
During adjudication, the verifier checks the supplied trace against the signed checkpoint the user filed, the final commitment, and every checkpoint on chain, so that changes to any covered record are detectable.

\subsection{User submission, adjudication and settlement}
\label{sec:design:sessions}
\label{sec:design:adjudication}
Before the session begins, the provider sends a blockchain transaction registering $s$, the user's payout address, and the public key matching the session's signing key. 
If the user finds a violated promise, they file a challenge with $s$, the violated promise, and, if they have one, the newest signed checkpoint from that session.
The contract checks its signature against the registered public key.
Here, users need to post a bond $b>0$.
This discourages users from filing unfounded claims in the hope that the provider fails to defend against them.
Malicious users can deliberately induce agents to violate promises and receive payment, e.g., via prompt injection \citep{debenedetti2025camel}; we view such behavior as in scope, as agents should follow policy regardless of user prompt injections. 

To settle a challenge, 
our implementation uses a trusted independent verifier. The provider sends $T$ and $\theta$ to the verifier, which checks them against the commitments and any signed checkpoint the user filed, evaluates $\phi(T;\theta)$, and submits a verdict.\footnote{In the absence of a trusted verifier, providers could produce a zero-knowledge proof of these checks, which the smart contract could verify without receiving the trace \citep{kosba2016hawk}.}
The contract authenticates the verifier and applies the settlement rules.
Algorithm~\ref{alg:claim-settlement} (Appendix~\ref{app:pseudocode}) specifies how the verifier checks an accepted claim and how the contract settles it.
If a violation is detected, the escrow pays the user the set-aside $p$ and returns $b$; otherwise, $p$ returns to the reserve and $b$ goes to the provider.
Each session--promise pair can receive at most one payout.
The provider must send its evidence within three days of filing.
If $T$ is unavailable by then or does not match the committed hashes or the user's signed checkpoint, the verifier rules against the provider and the user is paid as if a violation had been detected.
This design incentivizes providers to keep the record correct and available.
Appendix~\ref{app:design:settlement} lists every settlement outcome (Table~\ref{tab:outcomes}) and describes claim withdrawal, insufficient reserves, retirement, and the deadlines for filing claims.

\section{Implementation}
\label{sec:impl}
We implement the trusted-verifier design from Section~\ref{sec:design:adjudication} and integrate it with Hermes and OpenClaw, two agent harnesses.
Both integrations use the same SDK, trace store, verifier, and escrow contract.
Framework-specific changes capture their execution and permission events.

\mypara{Reference implementation}
\label{sec:impl:prototype}
Each harness adapter converts tool calls, permission decisions, and results into a common record format.
The shared Python SDK sends these records to a provider-hosted HTTP service, which stores the execution trace $T$ in SQLite.
For a checkpoint covering the first $n$ records, the SDK reads them back from the store and computes $h_n$. 
It submits $(s,n,h_n)$ to the escrow contract.
At the end of the session, it commits $H(T)$ for the complete stored trace.
When it opens a session, the SDK generates a signing key and registers its public key with the escrow contract. After every record, it signs $(s,n,h_n)$ and sends the signature to the user's client without waiting for the store.

A separate Python verifier service watches the escrow contract for challenges.
For each challenge, the provider signs $T$ and $\theta$ and sends them to the verifier, which reads the registered hashes and session commitments from the contract.
It recomputes their hashes, including every checkpoint, and checks that they match the commitments and any signed checkpoint the user filed.
The verifier evaluates $\phi(T;\theta)$ and submits its verdict to the %
escrow contract, which applies the settlement rules in Section~\ref{sec:design:adjudication}.
Appendix~\ref{app:implementation} documents the record format and deployment details.
To verify our implementation, we tested 16 challenges across 13 sessions on Base mainnet (Appendix~\ref{app:eval:practicality}). 

\mypara{Agent integration}
\label{sec:impl:integration}
We implement VAI with Hermes and OpenClaw, two prominent open-source agent harnesses~\citep{hermes2026interest,openclaw2026interest}. 
\textit{Hermes}
dispatches tool calls through a shared function in its Python harness.
Our adapter records each tool's arguments and their results. 
We also extend the harness' hooks to record each user's permission decision.
The adapter passes these records to the shared SDK.
We add recording hooks to \emph{OpenClaw}'s tool-execution and permission-handling paths, capturing tool arguments, approval decisions, and final results.\footnote{Because OpenClaw is written in TypeScript, a plugin forwards these events to a local Python process running our SDK.
OpenClaw thus shares storage, commitment, and adjudication code with Hermes.}

Both integrations use the same SDK, trace store, verifier, and code for record formats, hashing, and predicates.
These total 3,152 lines of Python, alongside a 489-line Solidity escrow contract.
The adapters are 217 lines for Hermes and 543 for OpenClaw; the harness patches add 180 and 305 lines and remove 37 and 3 (Table~\ref{tab:integration} in Appendix~\ref{app:implementation:integrations}).
The recording hooks and harness patches are reused across promises that use the same recorded information.
The required configuration or registration code is promise-dependent, including its predicates and/or recording requirements.

\section{Evaluation}
\label{sec:eval}
We evaluate VAI by 
assessing its policy violation coverage and operational costs.

\subsection{Policy violation coverage}
\label{sec:eval:effectiveness}

\mypara{Datasets} We evaluate coverage against standard benchmarks from the agent testing and security literature.
These benchmarks provide a \emph{task} the agent should complete, and natural-language instructions or policies that the agent must follow during execution.
We call a full sequence of interactions between the agent being tested and the benchmark environment a \emph{run} or \emph{session}.

\textit{(1) AgentDojo} evaluates agents on 97 distinct user tasks and 35 attacker goals, making up 949 task--attack combinations in banking, workspace, Slack, and travel environments while exposed to prompt injections~\citep{debenedetti2024agentdojo}. 
These attacks pursue specific goals, such as transferring money or sending private information to an attacker.
The benchmark separately checks whether the user's task and the attacker's goal succeed; we use successful attacks as the reference for evaluating VAI's detection coverage.

\textit{(2) $\tau^3$-bench} comprises 164 tasks (50 airline  and 114 retail), in which customer-service agents interact with simulated users~\citep{yao2024tau,sierra2026tau3}. 
The agents are instructed to follow policies such as obtaining confirmation before making changes,  checking eligibility for cancellations and returns, and following communication guidelines with customers.
However, the existing checks (and resulting dataset annotations) do not establish compliance with every applicable policy.
We therefore produce our own policy-compliance labels (see below).

\textit{(3) ClawsBench} evaluates agents on productivity tasks in simulated workplace services~\citep{li2026clawsbench}.
We use the 60-task \texttt{Standard60} cohort.
The benchmark checks whether agents complete the requested tasks under safety conditions. 
We produce additional policy-compliance labels. 

\mypara{Agents and instrumentation}
AgentDojo and $\tau^3$-bench take only an LLM as input (i.e., no harness) and manage its tool calls.
We ran two models, GLM-4.7-Flash and Qwen3-30B-A3B, on both benchmarks.
The benchmarks expose tools and record tool calls, which we used for VAI evaluation.
ClawsBench instead takes a full agent as input; we tested OpenClaw with GLM-5.2 as the underlying LLM.
We then evaluated VAI on the recorded OpenClaw traces and benchmark service logs.
Additionally, we also updated the harness in $\tau^3$-bench and ClawsBench to examine how more structured tool calls can enable better auditability. For instance, we added a tool in $\tau^3$-bench that lets the agent ask and obtain user permission as a structured data object instead of natural language.

\mypara{Policy construction}
VAI can detect and enforce only policies with objective criteria. 
To this end, we first manually identified the sub-policies for each benchmark.
For each objective sub-policy, we constructed a promise based on our predicate implementation.
For example for AgentDojo, we construct a promise that forbids the agent from sending a message containing the specific phishing link. 
For $\tau^3$-bench, we construct a promise that checks that an order is eligible before the agent cancels it.
For ClawsBench, we construct a promise that checks that a document's contents were read before external access was granted through Drive.
Once promises are constructed (details in Appendix~\ref{app:eval:setup}), we evaluate them on recorded agent traces.

\mypara{Policy-level dataset reannotation}
Existing annotations for $\tau^3$-bench and ClawsBench check high-level task outcomes and some safety requirements, but do not provide the policy-compliance labels we require to evaluate VAI.
We therefore re-annotated agent traces from  $\tau^3$-bench and ClawsBench to include policy-level violation information. 
Three LLMs independently labeled every run, and we use the majority label.
We use the resulting annotation as ground truth for our remaining experiments. 
To check these labels, two human annotators independently labeled a random sample of 30 runs from each benchmark.
The two human annotators then discussed and resolved their disagreements, and we use their resolved labels as the human baseline (details in Appendix~\ref{app:eval:labels}).
The majority label agrees with the human baseline on 27/30 runs on $\tau^3$-bench and 24/30 on ClawsBench.

On $\tau^3$-bench, our annotations identify at least one policy violation in 123 of 164 GLM-4.7-Flash runs and 151 of 163 Qwen3-30B-A3B runs.
On ClawsBench, 8 of 60 GLM-5.2 runs are labeled as containing a policy violation.
We believe this low occurrence may be due to using a relatively advanced LLM for OpenClaw on ClawsBench.

\mypara{Metrics}
For each successful attack in AgentDojo and each run labeled with at least one violation in $\tau^3$-bench and ClawsBench, we assign one of two categories:
    \textit{Detected with original harness:} VAI detects at least one violation from the records the benchmark's original harness logs.
    \textit{Detected with updated harness:} VAI detects at least one violation from the records the updated harness logs.

\begin{table}[t]
    \centering
    \small
    \setlength{\tabcolsep}{2.75pt}
    \caption{Coverage of violations across three benchmarks. AgentDojo uses successful attacks; $\tau^3$-bench and ClawsBench use our labels (violations). False alarms mean that VAI detected violations on runs without a violation.
    }
    \label{tab:eval:policy-coverage}
    \vspace{3pt}
    \begin{tabular*}{\linewidth}{@{\extracolsep{\fill}}llrrrrr@{}}
        \toprule
         & & & \multicolumn{2}{c}{Detected} & \multicolumn{2}{c@{}}{False alarms} \\
        \cmidrule(lr){4-5}\cmidrule(l){6-7}
         & & & Original & Updated & Original & Updated \\
        Benchmark & Model & Violations & harness & harness & harness & harness \\
        \midrule
        AgentDojo & GLM-4.7-Flash & 266 & 251 (94.4\%) & & 7/97 & \\
         & Qwen3-30B-A3B & 161 & 148 (91.9\%) & & 9/97 & \\
        $\tau^3$-bench & GLM-4.7-Flash & 123 & 20 (16.3\%) & 77 (62.6\%) & 0/41 & 0/41 \\
         & Qwen3-30B-A3B & 151 & 29 (19.2\%) & 87 (57.6\%) & 1/12 & 1/12 \\
        ClawsBench & GLM-5.2 & 8 & 3 (37.5\%) & 3 (37.5\%) & 3/51 & 5/51 \\
        \midrule
        All & & 709 & \textbf{451 (63.6\%)} & \textbf{566 (79.8\%)}\textsuperscript{a} & 20/298 (6.7\%) & 22/298 (7.4\%) \\
        \bottomrule
    \end{tabular*}

    \vspace{2pt}
    {\footnotesize\raggedright \textsuperscript{a}\,AgentDojo counted with its original-harness numbers.\par}
\end{table}

\subsection{Policy violation coverage: Main results}
Table~\ref{tab:eval:policy-coverage} summarizes the overall evaluation results for coverage of violation detection. It illustrates several key points. 

\mypara{Finding 1: Despite limited access to structured permission objects, VAI has a 63.6\% true positive rate (TPR) on our benchmarks}
Many agents do not explicitly emit structured data objects for events like user consent, permission changes, or security-sensitive actions. 
Even in the absence of such instrumentation, Table \ref{tab:eval:policy-coverage} shows that VAI has a 63.6\% TPR for policy violations overall.
\taubench and ClawsBench have substantially lower TPRs than AgentDojo (16-19\% for \taubench and 37.5\% for ClawsBench).
With only eight runs labeled as violations, ClawsBench provides limited evidence about detection coverage.
However, it informs us how to better record agent traces and clarify ambiguities in policy interpretations.
Specifically, on manual error analysis, we found that \taubench collects less structured information about agent interaction, and the OpenClaw agent (which we used for ClawsBench) barely emits any structured objects for security data at all.

\mypara{Finding 2: If agents are augmented to emit structured information about policy-relevant observations, the overall TPR jumps to 79.8\%}
Unsurprisingly, we find that most of our missed detections can be detected if agents were instrumented to log more information about actions the user authorized, including their scope and arguments---e.g., collecting user consent explicitly.
The resulting coverage is high across benchmarks and models, suggesting that VAI could be practically useful.
Table~\ref{tab:eval:records} in Appendix~\ref{app:eval:setup} lists the records the updated harness logs and the ones that are still missing.
In fact, we discuss how VAI could be applied to several real-world case studies in Appendix~\ref{sec:cases} 
where we argue that a properly-instrumented agent could have caught several of 
the high-profile failures introduced in Section \ref{sec:introduction}.

\textit{Recommendation:} 
Our finding highlights an important takeaway. Agents would be substantially more transparent 
and easier to monitor
if they emitted structured data objects corresponding to the information required to evaluate behavioral policies---not just for VAI, but more broadly. 
We call for the \textbf{co-design of agent architectures with their security policies} as a best practice for agent development: agent providers are best-suited to determine the information needed to evaluate their own agent policies, and should instrument their agents to explicitly emit such information. 

\mypara{Finding 3: Unstructured information and ambiguous policy exceptions can produce false positives}
Our analysis of false positives suggests that the main causes are unstructured information and, in some cases, ambiguity in policy interpretation.
For instance, in an AgentDojo task, the user asks the agent to email hotel details to their wife and supplies her address.
VAI flags the email because the address is absent from the structured recipient list used by the promise.
The address appears only in the user's natural-language instructions, which cannot be objectively parsed into that list.
On ClawsBench, VAI flags a violation in 5 of the 51 runs labeled `no violation' by the LLM committee.
In each case, our promise reads a rule more literally than the LLM committee did, such as requiring a content read before a share the user explicitly requested (explanation in Appendix~\ref{app:eval:claws-disagreements}).
Unstructured information accounts for 15/22 runs marked as false positives, and ambiguity in policy interpretation for 7/22.

\FloatBarrier
\subsection{Operating Cost}
\label{sec:eval:practicality}
Recording and committing a session's records requires under 20 ms per tool call, or under 80 ms with signed checkpoints, and under 1\textcent~on Base mainnet.
With OpenClaw, signed checkpoints add 49 ms per tool call, about 1.3\% of the 3.8 s the agent takes per tool call in our ClawsBench runs, including model generation (Appendix~\ref{app:eval:practicality}).
For users without signed checkpoints, more frequent on-chain checkpoints shorten the window in which records are vulnerable to alteration, but cost more in transaction fees (Table~\ref{tab:eval:policies}).
For a 10,000-record trace, the provider signs and sends the evidence in 0.4 seconds, and the verifier checks it and reaches a verdict in 0.3 seconds (Appendix~\ref{app:eval:practicality}). All in all, VAI's operating costs are low.

\begin{table}[!htbp]
    \centering
    \small
    \caption{Operating costs by checkpoint policy: Base mainnet fee per session and average record wait for Hermes and OpenClaw. Verifier processing time is a separate measurement for a 10,000-record trace (measured for the first and last policy); signing and delivering the trace adds 0.40 seconds. }
    \label{tab:eval:policies}
    \setlength{\tabcolsep}{3pt}
    \begin{tabular*}{\linewidth}{@{\extracolsep{\fill}}lrrrrr@{}}
        \toprule
        & \multicolumn{2}{c}{Fee per session (USD)} & \multicolumn{2}{c}{Record wait (s)} & Verifier \\
        \cmidrule(lr){2-3}\cmidrule(lr){4-5}
        Checkpoint policy & Hermes & OpenClaw & Hermes & OpenClaw & time (s) \\
        \midrule
        Every 10 records or 30 s (default) & 0.00512 & 0.00378 & 12.46 & 12.27 & 0.344 \\
        Every 30 records or 60 s & 0.00378 & 0.00244 & 11.25 & 10.91 & --- \\
        Final commitment only & 0.00086 & 0.00087 & 12.03 & 11.14 & 0.264 \\
        \bottomrule
    \end{tabular*}
\end{table}

\section{Related Work}

\mypara{Misbehavior prevention}
Recent work has explored strategies for preventing LLM agents from being manipulated into harmful or unintended behavior.
\textit{Progent} adds a programmable privilege-control layer and restricts LLM agents to only use permitted tools and actions~\citep{shi2025progent}. 
CaMeL separates the agent's control flow from untrusted data, so instructions injected into tool outputs cannot change that control flow, and CaMeL's security policies restrict how untrusted data reaches tool calls~\citep{debenedetti2025camel}.
DRIFT monitors agents' actions and filters instructions that conflict with the user's task~\citep{li2026drift}.
These methods constrain agent behavior during execution.
Our system complements them by adjudicating recorded violations after the fact
and enforcing payouts from provider deposits.

\mypara{Auditing and Accountability}
\citet{alqithami2025adaptive} traces responsibility among networked agents, detects harmful emergent norms such as collusion or hoarding, and recalibrates the agents' objectives or policies in near real time.
The difference is that its response and impact stay inside the agent system, whereas VAI pays the affected user from the provider's deposit once an independent verifier confirms a violation.
\citet{nian2026auditable} argue in a position paper that accountability requires trustworthy evidence to reconstruct actions, check compliance, and attribute responsibility.
VAI supplies that evidence through committed traces and signed checkpoints, similar to PeerReview's signed authenticators over a hash-chained log~\citep{haeberlen2007peerreview}, and adds the promise, the reserve, and the payout.
In concurrent work, a system called Recourse explores a similar idea, using smart contracts and provider deposits to pay for agent failures,
 but 
requiring providers to specify how to \emph{repair} damages once they occur%
~\citep{bindschaedler2026recourse}.
Unlike VAI, Recourse requires tool-specific recording and repair integration; by providing a tool-agnostic SDK, we were able to integrate VAI with prominent agents Hermes and OpenClaw.
Finally, while Recourse's empirical evaluation uses synthetic data and test environments with injected failures, we evaluate VAI on 
three established benchmark datasets; we also add new policy-level annotations of prominent benchmarks \taubench~\citep{yao2024tau,sierra2026tau3} and ClawsBench \citep{li2026clawsbench}.

\section{Conclusion and Limitations}
\label{sec:conclusion}
\label{sec:discussion}
\label{sec:discussion:future}

We introduced \name (VAI), a framework for financial recourse after agent failures, and implemented it end-to-end on Base mainnet.
While VAI serves as an early proof of concept for the idea of deterministic recourse for agent failures, it 
explicitly does not cover subjective policies, and even for policies deemed objective, we found both false negatives and (fewer) false positives, due to a combination of annotation challenges and unstructured or implicit tracing of agent actions. 
To address these challenges,  we need lightweight methods (both for existing and new agents) to augment agent trace logging for more transparent agent monitoring. 
VAI also only sees what a tool call records. For a general-purpose tool such as a shell, a promise can require the user's approval of the exact command, but it cannot check what that command did.
We also have not modeled price discovery for each promise's payout and reserve or evaluated the capital efficiency of reserves, which restaking could improve~\citep{eigenlabs2024security}.
A provider that finds a reliable way to induce a violation can also trigger it in its own sessions and drain the reserve back to its own wallet before actual users file claims. Users can check the reserve on chain before each session, but it can still be drained before they file. Since a drained reserve is public, this is a one-shot attack that costs the provider its reputation.
Resolving subjective promises and turning policy documents into promises also remain important open problems.

\subsubsection*{Acknowledgments}
We thank Shreyas Sharma for serving as the second annotator of the human baseline, and Takudzwa Tarutira for an early labeling of the \taubench and ClawsBench calibration runs. We gratefully acknowledge funding from the Initiative for CryptoCurrencies and Contracts (IC3) and their partners, as well as the Carnegie Mellon University Secure Blockchain Initiative (SBI) and its industry partners. 
This report describes research conducted by an undergraduate researcher supported through the REU program under National Science Foundation Grant No. CCF-2621730; it was also partially funded by NSF grant SaTC-2325477. 

\bibliographystyle{iclr2027_conference}
\bibliography{references}

\section*{AI Use Statement}
We used Claude Code and Codex to implement methods: they wrote parts of the prototype, the harness adapters, the evaluation pipeline, and the figure scripts under our direction, and we discussed design choices with them. All code and decisions were reviewed manually.
We used LLMs to write part of the code of the evaluation pipeline that categorized tasks labeled as violations.
LLMs were also used for data annotation, as described in Section~\ref{sec:eval:effectiveness} and Appendix~\ref{app:eval:labels}; our study design included a human validation pool to measure the quality of these annotations.
We did not use AI tools to develop theoretical models or mathematical claims; the remaining required-disclosure tasks do not apply to this work.
We used AI assistants to draft and polish text in every section.
We reviewed and heavily edited all writing.
Every sentence was edited by an author, every technical statement was checked against the code and the recorded results, and an author read every cited source.
We take responsibility for the final content of this work, including text, claims, and artifacts produced with the aid of AI tools.

\clearpage
\appendix
\renewcommand{\topfraction}{0.9}\renewcommand{\bottomfraction}{0.8}\renewcommand{\textfraction}{0.07}\renewcommand{\floatpagefraction}{0.85}
\FloatBarrier
\section{VAI Pseudocode}
\label{app:pseudocode}

\begin{algorithm}[!htbp]
    \caption{Claim verification and payment.
    }
    \label{alg:claim-settlement}
    \small\AlgSetup\setcounter{algln}{0}%
    \begin{itemize}[leftmargin=*,nosep,label={}]
    \item \textbf{Input:} a claim $c=(u,s,\pi,t_f,b,\kappa)$, filed by user $u$ against promise $\pi$ for session $s$ at time $t_f$ with bond $b$. $\kappa=(m,\hat h_m,\sigma)$ is the newest signed checkpoint the user holds, or $\bot$ if none.
    \item \textbf{Escrow state:} for $\pi$, the payout $p$, the reserve $R_t$, the hashes $h_\phi$ and $h_\theta$, and the queue $Q_\pi$ of unfunded claims, oldest first.
    For $s$, the public key $k_s$ of the session's signing key, each checkpoint $(n,h_n)$, where $h_n$ is the hash of the first $n$ records, and the final trace hash $h_s$.
    \item \textbf{Evidence:} $E=(T,\theta)$ from the provider, with $T$ the list of records it stored for $s$ and $\theta$ the parameters of $\pi$. We write $T_{1:n}$ for the first $n$ records of $T$.
    $T$ reproduces $\kappa$ if its first $m$ records chain to $\hat h_m$, or, for a pre-action checkpoint, record $m$ is the call $\kappa$ announces and that call ran (Appendix~\ref{app:implementation:signed}).
    \item \textbf{Output:} a verdict $v\in\{\texttt{violated},\texttt{satisfied}\}$ for $c$. A failed \kw{require} aborts the call and changes nothing.
    \end{itemize}

    \par\medskip
    \textbf{Escrow contract} (on chain), when a claim is filed\par\smallskip
    \begin{minipage}[t]{0.475\linewidth}\AlgSetup
    \Ln{0}{\Proc{File}($c$)}
    \Ln{1}{\kw{require} $u$ is the user of $s$}
    \Ln{1}{\kw{require} $(s,\pi)$ is unresolved and has no open claim}
    \Ln{1}{\kw{require} $b>0$}
    \Ln{1}{\kw{if} $\kappa\neq\bot$}
    \Ln{2}{\kw{require} $m>0$}
    \Ln{2}{\kw{require} $\sigma$ signs $(s,m,\hat h_m)$ under $k_s$}
    \Ln{1}{hold $b$}
    \Ln{1}{append $c$ to $Q_\pi$}
    \Ln{1}{\Proc{Fund}($\pi$)}
    \end{minipage}\hfill{\color{black!30}\vrule}\hfill
    \begin{minipage}[t]{0.475\linewidth}\AlgSetup
    \Ln{0}{\Proc{Fund}($\pi$)}
    \Ln{1}{\kw{while} $Q_\pi$ is not empty and $R_t\geq p$}
    \Ln{2}{$c'\gets$ the oldest claim in $Q_\pi$}
    \Ln{2}{remove $c'$ from $Q_\pi$}
    \Ln{2}{$R_t\gets R_t-p$}
    \Ln{2}{\kw{if} $c'$ is owed}
    \Ln{3}{credit $p$ to the user of $c'$}
    \Ln{2}{\kw{else}}
    \Ln{3}{mark $c'$ funded}
    \end{minipage}

    \par\medskip
    {\color{black!30}\hrule}
    \par\medskip
    \begin{minipage}[t]{0.475\linewidth}\AlgSetup
    \textbf{Verifier} (off chain)\par\smallskip
    \Ln{0}{\Proc{Verify}($c$)}
    \Ln{1}{wait for evidence $E=(T,\theta)$ for $c$ signed by the provider, or until now $>t_f+3$ days}
    \Ln{1}{\kw{if} no evidence arrived}
    \Ln{2}{\Proc{Settle}($c$, \texttt{violated}); \kw{return}}
    \Ln{1}{read $h_s$, every $(n,h_n)$, $h_\theta$, and $h_\phi$ from the escrow}
    \Ln{1}{\kw{if} $H(T)\neq h_s$, $H(\theta)\neq h_\theta$, or $H(T_{1:n})\neq h_n$ for any checkpoint}
    \Ln{2}{\Proc{Settle}($c$, \texttt{violated}); \kw{return}}
    \Ln{1}{\kw{if} $\kappa\neq\bot$ and $T$ does not reproduce $\kappa$}
    \Ln{2}{\Proc{Settle}($c$, \texttt{violated}); \kw{return}}
    \Ln{1}{$\phi\gets$ the catalog predicate whose hash is $h_\phi$}
    \Ln{1}{\kw{if} no such $\phi$ exists or $\phi(T;\theta)$ fails}
    \Ln{2}{\Proc{Settle}($c$, \texttt{violated})}
    \Ln{1}{\kw{else}}
    \Ln{2}{\Proc{Settle}($c$, $\phi(T;\theta)$)}
    \end{minipage}\hfill{\color{black!30}\vrule}\hfill
    \begin{minipage}[t]{0.475\linewidth}\AlgSetup
    \textbf{Escrow contract} (on chain)\par\smallskip
    \Ln{0}{\Proc{Settle}($c$, $v$)}
    \Ln{1}{\kw{require} $c$ is open}
    \Ln{1}{\kw{require} the sender is the verifier}
    \Ln{1}{\kw{if} $v=\texttt{violated}$}
    \Ln{2}{\kw{require} $(s,\pi)$ is unresolved}
    \Ln{2}{\kw{if} $c$ is funded}
    \Ln{3}{transfer $p+b$ to $u$}
    \Ln{2}{\kw{else}}
    \Ln{3}{transfer $b$ to $u$}
    \Ln{3}{mark $c$ owed, so \Proc{Fund} pays $p$ later}
    \Ln{1}{\kw{else} ($v=\texttt{satisfied}$)}
    \Ln{2}{\kw{require} the final hash $h_s$ is on chain}
    \Ln{2}{credit $b$ to the provider}
    \Ln{2}{\kw{if} $c$ is funded}
    \Ln{3}{$R_t\gets R_t+p$}
    \Ln{2}{\kw{else}}
    \Ln{3}{remove $c$ from $Q_\pi$}
    \Ln{1}{close $c$; mark $(s,\pi)$ resolved}
    \Ln{1}{\Proc{Fund}($\pi$)}
    \end{minipage}
\end{algorithm}

\FloatBarrier
\section{Design Details}
\label{app:design}

\subsection{Predicate catalog and specification}
\label{app:design:predicates}
We provide an example list of predicate classes that VAI can support (Table~\ref{tab:catalog}), which are implemented in our prototype and used in our evaluations. 
A provider selects a predicate and supplies the parameters $\theta$ that are specific to their harness.

\begin{table}[!htbp]
\centering
\caption{Implemented predicates, the evidence each needs, and the parameters the provider sets.}
\label{tab:catalog}
\small
\setlength{\tabcolsep}{4pt}
\renewcommand{\arraystretch}{1.15}
\begin{tabularx}{\linewidth}{@{}>{\raggedright\arraybackslash}p{0.17\linewidth} >{\raggedright\arraybackslash}X >{\raggedright\arraybackslash}X >{\raggedright\arraybackslash}X@{}}
\toprule
Predicate & Rule & Evidence needed & Parameters $\theta$ \\
\midrule
AAP-1:\newline Authorization & Each declared destructive call has prior permission covering it & Tool call and matching permission record & Tools that need permission \\
\addlinespace
AAP-2:\newline Scope & Each checked action meets its scope and state constraints & Tool arguments, recorded state, and permission changes & Checked tools, permitted scope, and state constraints \\
\addlinespace
AAP-3:\newline Payment & Each payment goes to an approved merchant and stays within the per-payment cap & Merchant and amount & Payment tools, merchant list, and amount cap \\
\addlinespace
AAP-4:\newline Recipients & Every recipient of a checked outbound action is permitted & Recipients and recorded permission changes & Outbound tools and permitted recipients \\
\addlinespace
AAP-5:\newline Action limits & Action counts, cumulative amounts, or repeated calls stay within their limits & Tool calls, arguments, results, and amounts & Checked tools, caps, and repetition conditions \\
\bottomrule
\end{tabularx}
\end{table}

\FloatBarrier

\subsection{Settlement outcomes}
\label{app:design:settlement}
In some cases, such as a missing trace or an incorrect hash, the user is also paid $p$ provided that $R_t \geq p$, so a payment does not by itself prove an agent violation. A failed transfer reverts the whole transaction and leaves the claim and balances unchanged.

\begingroup
    \small
    \setlength{\tabcolsep}{3pt}
    \setlength{\LTcapwidth}{\linewidth}
    \begin{longtable}{@{}>{\raggedright\arraybackslash}p{0.40\linewidth}>{\raggedright\arraybackslash}p{0.13\linewidth}>{\raggedright\arraybackslash}p{\dimexpr0.37\linewidth-6\tabcolsep\relax}>{\raggedright\arraybackslash}p{0.10\linewidth}@{}}
    \caption{Settlement outcomes of a claim and the minimum wait after filing for each.}
    \label{tab:outcomes} \\
        \toprule
        Condition & Who acts & Result ($R_t$, $b$, pair) & Wait \\
        \midrule
        \endfirsthead
        \multicolumn{4}{@{}l}{Table~\thetable{} (continued)} \\
        \toprule
        Condition & Who acts & Result ($R_t$, $b$, pair) & Wait \\
        \midrule
        \endhead
        \bottomrule
        \endfoot
        Evidence matches, $\phi$ violated & verifier & user gets the $p$ set aside from $R_t$ at filing, plus $b$; pair resolved & none \\
        Evidence matches, $\phi$ satisfied & verifier & set-aside $p$ returns to $R_t$, $b$ credited to provider, pair resolved & none \\
        No evidence from the provider, or no final commitment & verifier & as row 1 & $>3$ days \\
        Missing or inconsistent records, or a trace, checkpoint, or parameter hash mismatch & verifier & as row 1 & none \\
        Evidence matches, but the registered predicate is missing or fails while running & verifier & as row 1 & none \\
        The claim carries a signed checkpoint that the committed trace does not reproduce & verifier & as row 1 & none \\
        No verdict & user withdraws & set-aside $p$ returns to $R_t$ and funds the next waiting claim, $b$ returned, pair unresolved & $>10$ days \\
        $R_t<p$ at filing: the claim waits behind earlier claims & anyone adds funds, or earlier claims settle & $p$ set aside in filing order; a violation found first returns $b$ at once and pays $p$ when the funds reach the claim & none \\
    \end{longtable}
\endgroup

\paragraph{Funding and insufficient reserves}
\label{app:design:funding}
The provider funds a promise by depositing ETH in the escrow contract when registering it. Anyone may add funds while claims are open (see below). The provider picks the reserve size; the prototype does not determine the size of the reserve from expected claims or covered sessions.
After retiring the promise, the provider waits $\ell$ days (we chose $\ell=30$); if no claim is open, the provider can withdraw the remaining balance.
Filing a claim sets aside $p$ from $R_t$, oldest claim first (Section~\ref{sec:design:promises}).
While $R_t<p$, new claims are still accepted but wait in line until deposits or returned set-asides cover them. If a claim is found to be a violation before it is funded, the user gets $b$ back right away and $p$ once the funds reach the claim. Because the escrow contract is on a public blockchain, anyone can see a promise's status and the reserve left to back it.
Our contract tests cover these cases: claims are paid in filing order whatever the order of the verdicts, a provider's own claim cannot take the payout of a claim filed before it, and a claim filed against an empty reserve is paid once a deposit arrives.

\paragraph{Claim withdrawal}
\label{app:design:withdrawal}
A user can withdraw their open claim and recover $b$ more than ten days after filing in case the verifier stalls (Section~\ref{sec:gap}). If the claim was funded, its set-aside $p$ returns to $R_t$ and funds the next claim in line.
After a withdrawal, the pair stays unresolved and the user can file a new claim on it as long as the filing deadline below has not passed. %

\paragraph{Retirement and filing deadlines}
A provider can retire a promise, meaning it is no longer active for successive sessions. 
A session can be challenged until thirty days after the session's 
final commitment or thirty days after a promise is retired, whichever comes first.
Without a final commitment, only the retirement deadline applies.
Neither deadline cancels a claim that has already been filed.

\FloatBarrier

\subsection{Case Studies}
\label{sec:cases}
We examine nine reported real-world agent incidents to understand if and how VAI could have interacted with the underlying systems based on incident documentation. 
We identify which promises could have applied, what evidence they would have required, and which failures remain out of scope.

\paragraph{Destructive action without consent (AAP-1)}

\begin{itemize}
    \item \textit{Replit agent deleted a production database during a code freeze}~\citep{replit2025register}. 
If the provider had declared the database tool the agent used as destructive and recorded its execution without prior authorization, AAP-1 would identify that invocation as a violation.

\item \textit{OpenClaw purged emails of the Meta Superintelligence safety director}~\citep{yue2026fastcompany} is another case of destructive action without consent. The user attributed the unapproved deletions to the agent summarizing its conversation history and losing her instruction to wait for approval~\citep{yue2026compaction}. If the agent harness had recorded the user's consent structurally and the deletions happened through a structured tool call, AAP-1 could have identified each unapproved deletion as a violation.

If the deletions happened through an unstructured tool, such as a shell command, AAP-1 could flag only an unapproved shell call, not the emails it deleted.

\item 
\textit{A subagent running GPT-5.6 Sol expanded \texttt{\$HOME} incorrectly and ran \texttt{rm -rf} on the user's home directory}~\citep{shumer2026x}.  AAP-1 can require approval for an exact shell invocation. It cannot infer which files the command will delete or whether the user approved each deletion. A deletion-specific promise therefore needs to specify the relevant effects exposed by structured tools or additional instrumentation.
\end{itemize}

\paragraph{Action outside declared scope (AAP-2)}
\begin{itemize}
    \item 
\textit{Alibaba's ROME agent opened a reverse SSH tunnel and diverted GPU capacity to cryptocurrency mining; neither action was authorized by its task}~\citep{wang2025rome,sobrado2026forbes}. If the agent had opened the tunnel through a tool that records the destination host, an AAP-2 promise listing the allowed hosts could have flagged the connection. AAP-2 examines only the target(s) recorded by a given tool, so it would not catch a tunnel opened inside a shell command (i.e. without tool calling) or GPU usage unless they were invoked via tool-calling.

\item \textit{OpenAI agents used a package repository, provided for downloading software, to exchange messages across separate task executions 
~\citep{openai2026artifactorymessages}}.
While it is unclear whether the fault lies with the agents or the environment setup, a destination allowlist would not identify these writes because the repository was already permitted. An AAP-2 promise could instead restrict that repository to read operations. Evaluating it would require records identifying the repository, the operation, and its result. 
\end{itemize}

\paragraph{Payment outside the mandate (AAP-3)}
\begin{itemize}
    \item 
\textit{An attacker first sent Grok's wallet a Bankr membership NFT that elevated Grok's permissions, then had Grok decode a Morse-code message and pass the hidden transfer instruction to Bankrbot, a bot with direct access to a crypto wallet}~\citep{grok2026neuraltrust}. Bankrbot sent three billion DRB tokens, about \$150K, to the attacker. If the harness records the recipient and amount, AAP-3 can check them against the registered mandate. The verdict then depends on the payment, regardless of how the agent was persuaded. Changing the committed cap or allowlist requires registering a different promise.

\item \textit{
Lobstar Wilde sent all of its roughly 52 million tokens, about \$450K by its operator's estimate, to a stranger who had asked for 4 SOL (about \$320)}~\citep{pash2026lobster}. This incident was reported as an error. Given structured payment evidence, the AAP-3 predicate can check a cap or recipient restriction without establishing intent.
\end{itemize}

\paragraph{Recipient outside the allowlist (AAP-4)}
\begin{itemize}
    \item 
\textit{During training, an OpenAI agent uploaded a workbook to a public file-hosting service when its collaborators could not access the local file }~\citep{openai2026publicfilesharing}. The task called for local files, but the upload succeeded and produced a public download link. With upload destinations recorded, AAP-4 could identify transfers to hosts outside a registered allowlist. This requires recording the transfer itself; the current adapters do not extract its destination from arbitrary Python or shell code.
\end{itemize}

\paragraph{No fit}
\begin{itemize}
    \item 
\textit{An agent allegedly built on OpenClaw made a pull request to matplotlib; when it was rejected, the agent published a critical article, which the maintainer called a ``hit piece"~\citep{rathbun2026register}. }
A recorded publish call could expose the text, but VAI in its current form cannot determine whether a message is defamatory, as this is a subjective promise.
\end{itemize}

\FloatBarrier

\FloatBarrier
\section{Implementation Details}
\label{app:implementation}

\subsection{Recording and adjudication}
\label{app:implementation:recording}
Algorithm~\ref{alg:recording} shows how the SDK records a session and how the provider delivers evidence for a claim; the verifier and escrow contract then run Algorithm~\ref{alg:claim-settlement}.
The SDK numbers a session's records in order (\texttt{seq}) and sends them to a trace store, a database run by the provider. 
Records not yet sent are held in memory and can be lost if the host machine crashes. If the user holds a signed checkpoint covering a lost record, the stored trace can no longer reproduce it, so the verifier rules against the provider.
The store never overwrites a saved record, but the provider controls the database. 

\begin{algorithm}[!htbp]
    \caption{Recording a session and delivering evidence for a claim.}
    \label{alg:recording}
    \small\AlgSetup\setcounter{algln}{0}%
    \begin{itemize}[leftmargin=*,nosep,label={}]
    \item \textbf{Input:} a session $s$ covered by promise $\pi$, a trace store, the user's client, and the checkpoint thresholds of $k$ records and $r$ seconds.
    \item \textbf{Output:} signed checkpoints sent to the user's client, checkpoints $(n,h_n)$ and the final hash $h_s$ on the escrow, and signed evidence for each claim $c$ against $s$.
    Without a client, the SDK skips lines 2, 4, 7--9, and 12--15.
    \end{itemize}
    \par\medskip
    \begin{minipage}[t]{0.6\linewidth}\AlgSetup
    \textbf{SDK} (in the agent harness)\par\smallskip
    \Ln{0}{\Proc{Open}($s$)}
    \Ln{1}{generate a session key pair $(sk_s,k_s)$}
    \Ln{1}{open $s$ on the escrow with $k_s$; if this fails, run no tool call}
    \Ln{1}{$\hat h_0\gets H(\texttt{"VAI record chain v1"}\,\|\,s)$}
    \Ln{0}{\Proc{Append}(record)}
    \Ln{1}{append the record to $T$ as record $n$}
    \Ln{1}{$\hat h_n\gets H(\hat h_{n-1}\,\|\,H(T_n))$}
    \Ln{1}{sign $(s,n,\hat h_n)$ with $sk_s$ and send it to the client}
    \Ln{1}{\kw{for} each tool call announced but not yet recorded, sign its pre-action checkpoint again at count $n+1$}
    \Ln{1}{send unsent records to the store in order; keep the rest in memory if the store fails}
    \Ln{0}{\Proc{Record}(tool, args)}
    \Ln{1}{$d\gets H(\texttt{"VAI intent v1"}\,\|\,\text{tool, args, call id})$}
    \Ln{1}{sign the pre-action checkpoint $(s,n+1,H(\hat h_n\,\|\,d))$ and send it to the client}
    \Ln{1}{wait until the client has verified and stored every signed checkpoint so far}
    \Ln{1}{\kw{if} it does not, \Proc{Append} a record that the call did not run and \kw{return}}
    \Ln{1}{run the tool; keep its result or error (approvals go through \Proc{Append})}
    \Ln{1}{\Proc{Append}((seq, tool, args, result, time))}
    \Ln{0}{\Proc{Checkpoint}() \textit{after $k$ new records or $r$ seconds}}
    \Ln{1}{$n\gets$ number of records the store has saved}
    \Ln{1}{read $T_{1:n}$ back from the store; check it against earlier checkpoints}
    \Ln{1}{submit $(s,n,H(T_{1:n}))$ to the escrow}
    \Ln{1}{\kw{if} this fails, retry after $\min(r,5)$ seconds while the agent keeps running}
    \Ln{0}{\Proc{Close}()}
    \Ln{1}{\kw{require} no tool call is running and the store has saved every record}
    \Ln{1}{submit $h_s=H(T)$ of the stored trace once}
    \end{minipage}\hfill{\color{black!30}\vrule}\hfill
    \begin{minipage}[t]{0.36\linewidth}\AlgSetup
    \textbf{Provider} (off chain)\par\smallskip
    \Ln{0}{\Proc{Deliver}($c$)}
    \Ln{1}{\kw{require} $c$ is open and now $\leq t_f+3$ days}
    \Ln{1}{wait until $h_s$ is on the escrow}
    \Ln{1}{read $T$ and $\theta$ from the store}
    \Ln{1}{sign $(c,s,\pi,T,\theta)$ with the provider's registered key}
    \Ln{1}{send the evidence to the verifier, which keeps the first copy it receives}
    \end{minipage}
\end{algorithm}

\subsection{Signed checkpoints}
\label{app:implementation:signed}
The SDK hashes records one at a time: $\hat h_0=H(\texttt{"VAI record chain v1"}\,\|\,s)$ and $\hat h_n=H(\hat h_{n-1}\,\|\,H(T_n))$. After each record, it signs $(\text{chain id},\text{escrow address},s,n,\hat h_n)$ and sends it to the user's client. On-chain checkpoints hash the record list instead, so $\hat h_n\neq h_n$.

A tool call's record is written after the call returns, so a provider could run a call and never sign its record. To prevent this, before each tool call the SDK also signs a \emph{pre-action checkpoint} at count $n+1$, with $\hat h_{n+1}=H(\hat h_n\,\|\,H(\texttt{"VAI intent v1"}\,\|\,\text{tool, arguments, call id}))$, and runs the call only after the client has verified and stored it. If records are added before the call's own record, such as an approval, the SDK signs it again on top of them, so the call is always the next record.

The client keeps a checkpoint only if its signature matches the session's key, $n$ is one past the last one kept, and $\hat h_n$ extends the last kept hash. The contract checks the signature at filing (Algorithm~\ref{alg:claim-settlement}, line 7). The verifier checks that $T_{1:n}$ chains to $\hat h_n$, or that record $n$ is the announced call and it ran (line 27). Filing with a signed checkpoint costs 357k gas instead of 303k.

\subsection{Integration revisions and tested paths}
\label{app:implementation:integrations}
We integrated VAI with Hermes 0.16.0 and OpenClaw 2026.6.9~\citep{hermes2026snapshot,openclaw2026snapshot}.
Each integration consists of a patch, which adds the hooks we need to the harness's own source code, and an adapter, which is new code that passes each tool call and permission decision to our SDK.
Table~\ref{tab:integration} gives the size of each, and both are in the \texttt{integrations/} folder of our repository.
The line counts in Section~\ref{sec:impl:integration} are from the November 7, 2026 release, in which the SDK alone is 1,556 lines. They include comments and blank lines and exclude tests.

\begin{table}[!htbp]
\caption{Lines of code changed for each agent to integrate VAI, excluding tests and dependencies. For OpenClaw, the adapter count includes our TypeScript plugin and Python helper. 
}
\label{tab:integration}
\centering
\small
\setlength{\tabcolsep}{4pt}
\begin{tabularx}{\linewidth}{@{}>{\raggedright\arraybackslash}Xrr@{}}
\toprule
Integration changes & Hermes & OpenClaw \\
\midrule
Adapter code (lines) & 217 & 543 \\
Lines added/removed in the harness & $\textcolor[HTML]{1A7F37}{+180}/\textcolor[HTML]{CF222E}{-37}$ & $\textcolor[HTML]{1A7F37}{+305}/\textcolor[HTML]{CF222E}{-3}$ \\
Agent harness files changed & 7 & 6 \\
\bottomrule
\end{tabularx}
\end{table}

\FloatBarrier
\section{Evaluation Details}
\label{app:eval}

\subsection{Benchmark Settings and Promise Construction}
\label{app:eval:setup}
\paragraph{AgentDojo}
We use AgentDojo version 1.2.2. Nineteen allowlist promises (AAP-4)
restrict the destinations of payment, email, sharing, Slack, web, and calendar tools to those in structured fields of the pre-injection environment. The adapter also derives calendar-notification recipients from native tool behavior and prior calendar state.

\paragraph{$\tau^3$-bench}
We use Sierra's \texttt{tau2-bench} repository~\citep{sierra2026tau3} at revision \texttt{8ebb7499622f}, which identifies itself as $\tau^3$-bench. Each model runs every airline and retail task once through OpenRouter, with GPT-4.1-mini as the user simulator.
Each domain has one policy document that applies to all of its tasks. For each rule in that policy that can be checked from tool calls and their results, we write one promise, giving 21 airline and 9 retail promises. The same promises apply to every task in the domain. 

\paragraph{ClawsBench}
We use the Standard60 cohort of env0 version 0.2 (revision \texttt{d12ebf517ca7}) with BenchFlow version 0.6.5 (revision \texttt{737146cc60fb}). This 60-task snapshot differs from the 44-task set in the original paper~\citep{li2026clawsbench}. We run GLM-5.2 in OpenClaw 2026.7.1 with the supplied \texttt{assistant\_v1.md} policy as system context and without domain skills, automatic retries, or verifier feedback.
Similar to $\tau^3$-bench, we write a promise for each rule in ClawsBench's agent policy (\texttt{assistant\_v1.md}), which lists five safety rules (S1--S5) and five execution rules (E6--E10), that can be checked from tool calls and their results. 
\paragraph{Updated harness}
Additionally, when evaluating with the updated harness, we register more promises on $\tau^3$-bench based on data that become available through the updated harness, such as the user's approval of the exact call before each database write. On ClawsBench we register no new promise; each service now logs what it returned for every call and which account made the call, so the existing promises can be checked on every call. We also updated three sentences in the $\tau^3$-bench policy document to let the agent know what tools are now available and label the violations against the updated policy document for the updated harness.
Table~\ref{tab:eval:records} lists the records these promises read and the records VAI still lacks.
\begin{table}[!htbp]
    \centering
    \small
    \caption{Records read by the promises added with the updated harness, and records VAI still lacks.}
    \label{tab:eval:records}
    \begin{tabularx}{\linewidth}{@{}>{\raggedright\arraybackslash}Xlr>{\centering\arraybackslash}p{0.15\linewidth}>{\centering\arraybackslash}p{0.17\linewidth}@{}}
        \toprule
        Record & Benchmark & Runs\textsuperscript{b} & Deterministically extractable & Requires trusted semantic parsing \\
        \midrule
        The user's approval of the exact tool call & $\tau^3$-bench & 136 & \checkmark\textsuperscript{a} & \\
        The user's reason for a cancellation & $\tau^3$-bench & 14 & \checkmark\textsuperscript{a} & \\
        What each service call returned to the agent & ClawsBench & 0 & \checkmark & \\
        \addlinespace
        Whether what the agent told the user came from the user or the tools & $\tau^3$-bench & 81 & & \checkmark \\
        Whether a request was within what the agent can handle & $\tau^3$-bench & 17 & & \checkmark \\
        A recipient the user named only in their message & AgentDojo & 15 & & \checkmark \\
        \bottomrule
    \end{tabularx}

    \vspace{2pt}
    {\footnotesize\raggedright \textsuperscript{a}\,Recorded by the approval step we added to the $\tau^3$-bench harness.\\
    \textsuperscript{b}\,Violation runs detected by a promise that reads the record (first three rows), violation runs VAI misses (next two), and false alarms (last row).\par}
\end{table}

\subsection{Dataset Annotation}
\label{app:eval:labels}
The three LLMs are Fable 5, GPT 5.6 Sol, and Kimi K3.
Table~\ref{tab:eval:label-agreement} reports the agreement between the two human annotators before they resolved their disagreements, and the agreement between the human baseline and the LLM committee.
\begin{table}[!htbp]
    \centering
    \small
    \caption{Agreement on the 30 sampled runs from each benchmark, with Cohen's $\kappa$ in parentheses.}
    \label{tab:eval:label-agreement}
    \begin{tabular}{@{}lll@{}}
        \toprule
        Comparison & $\tau^3$-bench & ClawsBench \\
        \midrule
        Annotator 1 and Annotator 2 & 26/30 ($\kappa=0.64$) & 27/30 ($\kappa=0.52$) \\
        Human baseline and LLM committee & 27/30 ($\kappa=0.71$) & 24/30 ($\kappa=0.14$)\textsuperscript{a} \\
        \bottomrule
    \end{tabular}

    \vspace{2pt}
    {\footnotesize\raggedright \textsuperscript{a}\,$\kappa$ is low because the human baseline has only 5 violations on ClawsBench and the LLM committee found 1 of them.\par}
\end{table}

In cases where a violation was detected, the LLMs did not always agree on \emph{which} policy had been violated. 
In these cases, we recorded the policies that at least two LLMs marked as violated.
In $\tau^3$-bench, we measured the number of runs where VAI detected a policy violation; among these runs, we counted the number of times the violated policy (as detected by VAI) matched at least one of the violated policies marked by the LLMs. This agreement rate was 17/20 for GLM and 27/29 for Qwen.

\subsection{False Positives}
\label{app:eval:claws-disagreements}
Table~\ref{tab:eval:policy-coverage} counts 22 runs with false alarms, 16 in AgentDojo, one in $\tau^3$-bench, and five in ClawsBench with the updated harness. With the original harness, ClawsBench has three, for 20 in total.
We systematically analyze each of these errors below.

\paragraph{Unstructured information (15 runs)}
In all 7/7 GLM runs marked as false positives and 8/9 Qwen runs marked as false positives on AgentDojo, the user authorizes the recipient only in free text, which the allowlist does not use.
This comprises the bulk of our false positives.
In two of the Qwen runs, VAI also flags a real agent error: in banking task 15 the agent sends money to the placeholder \texttt{UNKNOWN\_IBAN}, and in Slack task 20 it sends an invitation with an invented address, \texttt{dora@example.com}, which the simulator ignores.

\paragraph{Ambiguous policy (7 runs)}
In AgentDojo Slack task 4, the agent posts to the \texttt{/hobbies} page of the requested website, which is authorized only if permission covers the whole site.

In $\tau^3$-bench airline task 18, the user downgrades five reservations from business to economy. Economy comes with fewer free checked bags, so our registered promises expect the bags above the new allowance to become paid. The policy does not say what happens to bags already on a reservation when its cabin changes, and the benchmark's own reference solution changes only the cabin. The labelers did not count it as a violation.

The five ClawsBench runs involve three rules.
In \texttt{auth-delegated-access-sharing}, the user names a file and asks the agent to share it with a colleague outside the company. The agent finds the file by name and shares it without opening it. Rule S4 asks the agent to read content before sharing it externally, but here the user already chose the file.
In \texttt{slack-channel-audit-and-notify}, the agent archives two small channels after reading their member counts from the channel list and their recent messages. Rule S3 asks for verification before a destructive action. Our promise accepts only the per-channel information request as verification, so it fires although the agent did verify.
The other three runs exceed the cap of three identical requests. Rule E8 tells the agent to stop and change strategy after more than three identical calls, which targets an agent stuck in a loop, while our promise counts every identical request. In \texttt{auth-token-expiry-recovery}, the agent rereads the label list while it creates labels and tags messages. In \texttt{multi-rebalance-on-call-rotation}, it rereads a document several times around its one edit. In \texttt{gdrive-bulk-reshare-confidential-doc}, it lists the same Drive files four times with no change in between, which exceeds the cap as written. The LLM committee counted none of the five as a violation. Two of the five are in the human sample, and the human baseline counts \texttt{multi-rebalance-on-call-rotation} as a violation of E8.

\FloatBarrier
\subsection{Operating-Cost Measurements}
\label{app:eval:practicality}
We measure what VAI costs to run in three parts: the time recording adds to each tool call, the fees and waits of on-chain checkpoints, and the verifier's time and memory. In the first two, an agent session writes 1\,KiB files through the harness's own file-write tool under two promises: an authorization promise (AAP-1) that requires approval for each write, and a scope promise (AAP-2) that limits writes to a workspace folder. On mainnet, each promise has a payout of $10^{10}$ wei and each claim a bond of $10^{10}$ wei. The verifier measurement uses traces built from the recorded writes.

\paragraph{Mainnet challenges}
To test and measure the cost of the claim flow end to end, we filed 16 challenges, mimicking a user, on Base mainnet from a separate user wallet, against 13 Hermes sessions.
Six challenges target three approved writes under both the authorization (AAP-1) and the scope (AAP-2) promise, and all six settle as no violation. Eight target a write outside the scope (three), a write without the required approval (three), evidence altered after the final commitment (one), and a final trace that contradicts a checkpoint (one); all eight settle as violations. The last two target a write outside the scope that the provider deletes from its records before the traces are committed onchain. Without the user's signed checkpoint, the committed trace no longer shows the write, and the challenge settles as no violation. With the signed checkpoint, the committed trace does not reproduce it, and the challenge settles as a violation. The writes go through Hermes's own \texttt{write\_file} tool and approval hooks, with scripted approval answers. Filing a challenge cost 1.84\,$\mu$ETH in fees on average and its verdict 0.61\,$\mu$ETH, about 0.5\textcent{} and 0.2\textcent{}.

\paragraph{Runtime protocol}
Table~\ref{tab:eval:recording-cost} reports how long a session of 100 file writes of 1\,KiB each takes, with recording disabled, with recording to the trace store, with recording plus commitments to a local chain, and with all of that plus signed checkpoints.
The difference from the first row is the time VAI adds. The times cover the tool calls and the session's final commitment; they exclude model generation and waiting for human approval. Each configuration has one discarded warmup and 30 measured repetitions in shuffled order.
For comparison, in the ClawsBench runs on the updated harness, 57 runs made 1,255 tool calls in 4,782 s of agent execution, or 3.8 s per tool call including model generation; we exclude two runs that timed out and one that made no tool call.

\begin{table}[!htbp]
    \centering
    \small
    \setlength{\tabcolsep}{3pt}
    \caption{Mean local runtime of 100 writes, including finalization and excluding setup, with 95\% confidence intervals for added time in brackets and Anvil as the local Ethereum test chain.}
    \label{tab:eval:recording-cost}
    \begin{tabularx}{\linewidth}{@{}>{\raggedright\arraybackslash}Xrr>{\centering\arraybackslash}p{0.17\linewidth}>{\centering\arraybackslash}p{0.22\linewidth}@{}}
        \toprule
        & \multicolumn{2}{c}{Mean time (s)} & \multicolumn{2}{c}{Added time (s)} \\
        \cmidrule(lr){2-3}\cmidrule(l){4-5}
        Configuration & Hermes & OpenClaw & Hermes & OpenClaw \\
        \midrule
        Recording disabled & 5.91 & 0.021 & --- & --- \\
        Recording and HTTP store & 6.70 & 0.538 & 0.79\newline [0.71, 0.87] & 0.517\newline [0.508, 0.527] \\
        \addlinespace
        Recording, store, and local chain & 7.74 & 1.088 & 1.83\newline [1.73, 1.96] & 1.067\newline [1.017, 1.123] \\
        \addlinespace
        Recording, store, local chain, and signed checkpoints & 13.90 & 5.985 & 7.99\newline [7.86, 8.13] & 5.963\newline [5.843, 6.077] \\
        \bottomrule
    \end{tabularx}
\end{table}

\paragraph{Checkpoint protocol}
Each of the three checkpoint policies in Table~\ref{tab:eval:policies} runs three mainnet sessions of 20 writes per harness. In each session, the agent makes two writes, idles for 35 seconds so that a policy's time limit can trigger a checkpoint, and then makes the remaining 18 writes 0.25 seconds apart.
We measure how long each record waits before a commitment covers it, from the time the store saves the record to the first receipt the client sees for a commitment that includes it.
These receipts are Base preconfirmations, so the waits do not measure block inclusion or Ethereum finality. The SDK submits checkpoints serially, so records produced during a submission wait for a later commitment. Under the ten-record, thirty-record, and final-only policies, the mean per-run maximum Hermes wait is 33.9, 41.4, and 46.8 seconds. The ten-record policy therefore has the shortest maximum wait but a longer average wait than the thirty-record policy. Session fees cover the checkpoints and final commitment, including execution and Ethereum (L1) data fees, and exclude promise registration and session opening.

\paragraph{Verifier protocol}
Table~\ref{tab:eval:verifier-cost} reports how long the verifier takes to check a trace and reach a verdict, and its peak memory, for traces of 100 to 10,000 records with and without on-chain checkpoints. Each claim carries a signed checkpoint covering the whole trace, the most work that check can take.
Before verification, the provider signs the trace and sends it to the verifier over local HTTP. For a 10,000-record trace with 1,000 checkpoints, signing takes 0.069 seconds and delivery 0.334 seconds, averaged over five runs.

\begin{table}[!htbp]
    \centering
    \small
    \caption{Mean verifier processing time over five fresh processes, excluding signing, delivery, blockchain requests, and settlement, with prefix checks and the signed-checkpoint check counted in the total and peak memory including imported libraries.}
    \label{tab:eval:verifier-cost}
    \begin{tabular}{@{}rrrrr@{}}
        \toprule
        Records & Checkpoints & Total (s) & Prefix checks (s) & Peak memory (MiB) \\
        \midrule
        100 & 0 & 0.0045 & $<0.0001$ & 80.2 \\
        100 & 10 & 0.0050 & 0.0008 & 80.2 \\
        1,000 & 0 & 0.0271 & $<0.0001$ & 87.0 \\
        1,000 & 100 & 0.0349 & 0.0076 & 86.4 \\
        10,000 & 0 & 0.2635 & 0.0002 & 160.8 \\
        10,000 & 1,000 & 0.3436 & 0.0772 & 161.2 \\
        \bottomrule
    \end{tabular}
\end{table}

\end{document}